\documentclass{article}
\usepackage{amsmath, amssymb, amsfonts}
\title{On a star with conformally flat geometry inside} 
\author{Hristu Culetu\footnote{electronic address: hculetu@yahoo.com} \\ Constanta, Romania}
\begin{document}
\numberwithin{equation}{section}
\pagenumbering{arabic}
\maketitle
\newcommand{\fv}{\boldsymbol{f}}
\newcommand{\tv}{\boldsymbol{t}}
\newcommand{\gv}{\boldsymbol{g}}
\newcommand{\OV}{\boldsymbol{O}}
\newcommand{\wv}{\boldsymbol{w}}
\newcommand{\WV}{\boldsymbol{W}}
\newcommand{\NV}{\boldsymbol{N}}
\newcommand{\hv}{\boldsymbol{h}}
\newcommand{\yv}{\boldsymbol{y}}
\newcommand{\RE}{\textrm{Re}}
\newcommand{\IM}{\textrm{Im}}
\newcommand{\rot}{\textrm{rot}}
\newcommand{\dv}{\boldsymbol{d}}
\newcommand{\grad}{\textrm{grad}}
\newcommand{\Tr}{\textrm{Tr}}
\newcommand{\ua}{\uparrow}
\newcommand{\da}{\downarrow}
\newcommand{\ct}{\textrm{const}}
\newcommand{\xv}{\boldsymbol{x}}
\newcommand{\mv}{\boldsymbol{m}}
\newcommand{\rv}{\boldsymbol{r}}
\newcommand{\kv}{\boldsymbol{k}}
\newcommand{\VE}{\boldsymbol{V}}
\newcommand{\sv}{\boldsymbol{s}}
\newcommand{\RV}{\boldsymbol{R}}
\newcommand{\pv}{\boldsymbol{p}}
\newcommand{\PV}{\boldsymbol{P}}
\newcommand{\EV}{\boldsymbol{E}}
\newcommand{\DV}{\boldsymbol{D}}
\newcommand{\BV}{\boldsymbol{B}}
\newcommand{\HV}{\boldsymbol{H}}
\newcommand{\MV}{\boldsymbol{M}}
\newcommand{\be}{\begin{equation}}
\newcommand{\ee}{\end{equation}}
\newcommand{\ba}{\begin{eqnarray}}
\newcommand{\ea}{\end{eqnarray}}
\newcommand{\bq}{\begin{eqnarray*}}
\newcommand{\eq}{\end{eqnarray*}}
\newcommand{\pa}{\partial}
\newcommand{\f}{\frac}
\newcommand{\FV}{\boldsymbol{F}}
\newcommand{\ve}{\boldsymbol{v}}
\newcommand{\AV}{\boldsymbol{A}}
\newcommand{\jv}{\boldsymbol{j}}
\newcommand{\LV}{\boldsymbol{L}}
\newcommand{\SV}{\boldsymbol{S}}
\newcommand{\av}{\boldsymbol{a}}
\newcommand{\qv}{\boldsymbol{q}}
\newcommand{\QV}{\boldsymbol{Q}}
\newcommand{\ev}{\boldsymbol{e}}
\newcommand{\uv}{\boldsymbol{u}}
\newcommand{\KV}{\boldsymbol{K}}
\newcommand{\ro}{\boldsymbol{\rho}}
\newcommand{\si}{\boldsymbol{\sigma}}
\newcommand{\thv}{\boldsymbol{\theta}}
\newcommand{\bv}{\boldsymbol{b}}
\newcommand{\JV}{\boldsymbol{J}}
\newcommand{\nv}{\boldsymbol{n}}
\newcommand{\lv}{\boldsymbol{l}}
\newcommand{\om}{\boldsymbol{\omega}}
\newcommand{\Om}{\boldsymbol{\Omega}}
\newcommand{\Piv}{\boldsymbol{\Pi}}
\newcommand{\UV}{\boldsymbol{U}}
\newcommand{\iv}{\boldsymbol{i}}
\newcommand{\nuv}{\boldsymbol{\nu}}
\newcommand{\muv}{\boldsymbol{\mu}}
\newcommand{\lm}{\boldsymbol{\lambda}}
\newcommand{\Lm}{\boldsymbol{\Lambda}}
\newcommand{\opsi}{\overline{\psi}}
\renewcommand{\tan}{\textrm{tg}}
\renewcommand{\cot}{\textrm{ctg}}
\renewcommand{\sinh}{\textrm{sh}}
\renewcommand{\cosh}{\textrm{ch}}
\renewcommand{\tanh}{\textrm{th}}
\renewcommand{\coth}{\textrm{cth}}

\begin{abstract}
 The properties of a star with constant positive energy density inside (as for the Schwarzschild interior geometry)  and a negative pressure are investigated, using a static conformally flat spacetime. Because of the negative pressure, the gravitational field inside is repulsive. Ricci and Kretschmann curvature invariants are finite. The energy conditions for the stress tensor of the perfect fluid are satisfied, excepting the strong energy condition which is not obeyed for $r<R/\sqrt{2}$, where $R$ is the radius of the object. The Komar mass is calculated and discussed.
 \end{abstract}
 
 \section{Introduction}
 By extending the concept of Bose–Einstein condensation to gravitational systems, Mazur and Mottola \cite{MM} constructed a static, spherically symmetric solution to Einstein’s equations, characterized by an interior de Sitter (deS) region with the equation of state $p = -\rho$, representing the gravitational vacuum condensate and an exterior Schwarzschild geometry of total mass $m$. There is a boundary between them, with a small thickness that replaces the both Schwarzschild and de Sitter classical horizons, resulting a compact object with no horizons. It is worth noting that a spatially homogeneous Bose-Einstein condensate (BEC) couples to Einstein’s equations in the same way as an effective cosmological term (deS equation of state). A similar vacuum negative pressure has been introduced by Li et al. \cite{LLZ} on their study about the Bag model on hadrons and the quark stars, the negative pressure $p$ playing the role of the bag constant.

Quantum corrections to Einstein's equation could be relevant at macroscopic scales and near event horizons. These arise from the conformal scalar degrees of freedom in the effective field theory of gravity generated by the trace anomaly of massless quantum
fields in curved space \cite{EM}.  At event horizons of black holes (BHs) the conformal anomaly degrees of freedom can have macroscopically large effects on the geometry, potentially removing the classical event horizon. The cosmological term becomes a dynamical condensate, whose value depends on boundary conditions near the horizon. In the conformal phase where the anomaly induced fluctuations dominate, the effective cosmological “constant” becomes a running coupling. By taking a positive value in the interior of a fully collapsed star, the effective cosmological term removes any singularity, replacing it with a smooth dark energy interior \cite{EM, MM}.  The apparent existence of cosmological dark energy, which is causing the expansion of the universe to accelerate, has the same equation of state as that of the quantum vacuum itself. The conformal trace anomaly of massless fields in curved space becomes large (formally infinite) for generic quantum states at both the Schwarzschild BHs and deS static horizons \cite{MV}. 

 Recently, Melella and Reyes \cite{MR}  found the interior solution for a static, spherically symmetric perfect fluid star backreacted
by QFT in four dimensions, with a constant energy density. The authors present the first exact self-consistent solution of a star that takes into account the backreaction effects of QFT. The source stress tensor is described by a perfect fluid. Instead of providing an equation of state relating the energy density $\rho$ and the pressure $p$, Mellela and Reyes impose some symmetry properties, e.g. they use conformally flat geometry in the star interior, when the Weyl tensor $W_{abcd} = 0$. This provides the extra condition that is necessary to find $\rho$ and $p$, and from here the equation of state. That way leads Mellela and Reyes to the conformally flat Schwarzschild interior solution, with the well known divergence of the $p(r)$, as Buchdahl showed many years ago\cite{HB}. Moreover, Buchdahl argued that the Schwarzschild interior solution is the only static, conformally flat solution for the star interior, with positive pressure and energy density. He also found the manifestly conformally flat form of the Schwarzschild interior, which is, however, time dependent \cite{HB}.

 In the semiclassical treatment, the authors of \cite{MR} take into account the conformal trace anomaly induced by the curvature (the fact that, in the quantum treatment, the trace of the energy-momentum of a massless quantum field is nonzero). 

Our goal in this paper is to find a static, conformally flat solution for the interior of a star, having a perfect fluid with constant energy density as the source of curvature in the Einstein equations. However, to achieve that purpose we have to relax the condition that $p\geq 0$ inside the star. We remember that one obtains a negative pressure for the Schwarzschild interior metric too, when the Buchdahl condition $R>9m/4$ is not obeyed. In addition, we get a regular metric with no horizons and finite curvature invariants. Moreover, the source stress tensor fulfills the weak, null and dominant energy conditions but the strong one is satisfied only for $r>R\sqrt{2}/2$. 

 The geometric units $c = G = 1$ and the positive signature $+2$ are used.  

\section{Conformally flat metric}
As we already specified in the Introduction, one looks for a general static conformally flat geometry with a conformal factor $f(r)>0$ depending only on the radial coordinate. Let us take the metric in the form
 \begin{equation}
ds^{2} = e^{2f(r)}(-dt^{2} + dr^{2} + r^{2}(d\theta^{2} + sin^{2}\theta d\phi^{2})),~~~~r\leq R,
 \label{2.1}
 \end{equation}
where $R$ is the radius of the spherically symmetric object (a star for example). The stress tensor inside the object is supposed to be consonant with a perfect fluid
\begin{equation}
T_{ab} = (p + \rho )u_{a}u_{b} + p g_{ab} ,
 \label{2.2}
 \end{equation}
where $a, b$ take the values $t, r, \theta, \phi$, $p$ and $\rho$ are the pressure and, respectively the energy density of the fluid. We choose the velocity vector field $u^{a}$ to represent a static observer such that $u^{a} = (e^{-f(r)}, 0, 0, 0)$, with $u^{a}u_{a} = -1$. The covariant acceleration $a^{b}$  of that static observer is given by
\begin{equation}
a^{b} = u^{b}\nabla_{b}u^{a} = (f'e^{-2f}, 0, 0, 0),~~~~f' = df/dr .
 \label{2.3}
 \end{equation}
Our next purpose is to find the expression of $f(r)$ from the proposed line-element (2.1) and Einstein's equations $G_{ab} = 8\pi T_{ab}$, where $G_{ab}$ is the Einstein tensor and $T_{ab}$ is given by (2.2). Inserting Eqs. (2.1) and (2.2) in gravitational equations one obtains
\begin{equation}
\begin{split}
G^{t}_{~t} = \left(2f'' + f'^{2} + \frac{4}{r}f'\right)e^{-2f} = -8\pi \rho,~~~G^{r}_{~r} = \left(3f'^{2}  + \frac{4}{r}f'\right)e^{-2f} = 8\pi p\\
G^{\theta}_{~\theta} = G^{\phi}_{~\phi} = \frac{1}{r}\left(2f' + rf'^{2} + 2rf''\right)e^{-2f} = 8\pi p .
\end{split}
 \label{2.4}
 \end{equation}
From $G^{r}_{~r} = G^{\theta}_{~\theta}$ we have 
\begin{equation}
2rf'' - 2rf'^{2} - 2f' = 0.
 \label{2.5}
 \end{equation}
This simple differential equation gives us $f(r)$ in terms of two constants of integration $\alpha$ and $\beta$
\begin{equation}
f(r) = log\frac{\alpha}{r^{2} + \beta},~~~~\alpha , \beta >0.
 \label{2.6}
 \end{equation}
Once $f(r)$ is determined, we get the energy density from the expression of $G^{t}_{t}$. Hence
\begin{equation}
8\pi \rho = \frac{12\beta}{\alpha^{2}},
 \label{2.7}
 \end{equation}
which shows that $\rho$ = const. inside the star. Noting that the same property is valid for the interior Schwarzschild geometry. The equation $m = (4/3) \pi R^{3}\rho$ will hold. This gives a relation between $\alpha$ and $\beta$ from (2.7)
\begin{equation}
\frac{2\beta}{\alpha^{2}} = \frac{m}{R^{3}}
 \label{2.8}
 \end{equation}
For the pressure of the fluid we get
\begin{equation}
8\pi p = \frac{4}{\alpha^{2}}(r^{2} - 2\beta ).
 \label{2.9}
 \end{equation}
One imposes the restriction $p(R) = 0$ on the object surface, supposing that it is isolated from other bodies. It yields $\beta = R^{2}/2$, and from (2.8), $\alpha = R^{2}\sqrt{R/m}$. We have finally
 \begin{equation}
8\pi p = \frac{4m}{R^{5}}(r^{2} - R^{2}).
 \label{2.10}
 \end{equation}
It is worth observing that $p(0) = -m/(2\pi R^{3})$. We see that the pressure is always negative, having a minimum at the origin. It is a monotonic function of $r$. As we noticed in the Introduction, even for the Schwarzschild interior geometry the pressure becomes negative for $r<9m/4$. In addition, it could be divergent in the same region. Even though the pressure is negative in our situation, it is however finite. Because of $p<0$, the gravitational field inside is repulsive, like for the static deS geometry. That results from the expression of the radial acceleration (2.3) which, with the help of $u^{t} = (r^{2} + \beta)/\alpha $ becomes
\begin{equation}
a^{r} = -\frac{mr}{R^{5}}(2r^{2} + R^{2}),
 \label{2.11}
 \end{equation}
with $a^{r}(0) = 0$ and $a^{r}(R) = -\frac{3m}{R^{2}}$. The radial acceleration is a monotonically decreasing negative function for any $r\in [0, R]$. One notices that $|a^{r}(R)|$ is close to the Newtonian value $Gm/R^{2}$. We have to keep in mind that $a^{r}$ is the acceleration needed to maintain a test particle at a constant location. 

It is worth finding the invariant acceleration of our static observer. It is given by
\begin{equation}
A \equiv \sqrt{a^{b}a_{b}} = \frac{2mr}{R^{3}}\sqrt{\frac{R}{m}} \propto r ,
 \label{2.12}
 \end{equation}
with $A(R) = (2m/R^{2})\sqrt{R/m}$. If we apply this for a star like our Sun, having the mass $2\cdot 10^{33}g$ and radius $\approx 7\cdot 10^{10}$cm, one obtains $A(R) \approx 5\cdot 10^{5}m/s^{2}$, much larger than $a\approx M_{\odot}/R_{\odot}^{2} = 274 ~m/s^{2}$. The difference comes from the factor $\sqrt{R/m}$, which is much greater than unity. 

Concerning the curvature invariants, one finds they are regular for any $0\leq r\leq R$. The scalar curvature and the Kretschmann invariant are given by
\begin{equation}
R^a_{~a} = \frac{12}{\alpha^{2}}(3\beta - r^{2}),~~~~K = \frac{48}{\alpha^{4}} \left[(r^{2} - \beta)^{2} + 4\beta^{2}\right].
 \label{2.13}
 \end{equation}

Let us investigate now whether the energy conditions for $T^{a}_{~b}$ are fulfilled. It is clear that $\ro >0,~ \rho + p>0$ and $\ro >|p|$. Hence, the null (NEC), weak (WEC) and dominant energy condition (DEC) are satisfied. However, the strong energy condition (SEC) is not satisfied for any $r$ because $\ro + 3p<0$ for $r\sqrt{2}<R$.

 The problem of the matching conditions at the boundary $r = R$ between the interior geometry and the empty Schwarzschild exterior is not completely solved only with the restriction $p(R) = 0$ at the interface. We need for that a more elaborate recipe. Mazur and Mottola \cite{MM}, for to satisfy the junction conditions and to assure stability, investigated a phenomenological model consisting of three regions: a deS interior metric, followed by a thin shell and then a Schwarzschild exterior. We let the matching problem for a future investigation.

\section{Komar mass}
Let us investigate now the Komar mass associated to the geometry (2.1), with $f(r)$ from (2.6)
 \begin{equation}
ds^{2} = \frac{\alpha^{2}}{(r^{2} + \beta )^{2}} \left(-dt^{2} + dr^{2} + r^{2}(d\theta^{2} + sin^{2}\theta d\phi^{2})\right).
 \label{3.1}
 \end{equation}
For that purpose we use the well-known expression \cite{TP}
 \begin{equation}
W_{K} = 2\int_{V}\left(T_{ab} - \frac{1}{2}g_{ab}T^{c}_{~c}\right)u^{a}u^{b}N\sqrt{h}~d^{3}x,
 \label{3.2}
 \end{equation}
where $N = \sqrt{-g_{tt}}$ is the lapse function and $h$ is the determinant of the spatial 3-metric, $h_{ab} = g_{ab} + u_{a}u_{b}$, namely $h = \alpha^{3}r^{2}sin\theta /(r^{2} + \beta)^{3}$. One finds that
 \begin{equation}
W_{K}(r) = \frac{48 R^5}{m}\int_{0}^{r}\frac{y^{2}(2y^{2} - R^{2})}{(2y^{2} + R^{2})^{4}}dy.
 \label{3.3}
 \end{equation}
One observes that the fraction inside integral is negative for $y < R/\sqrt{2}$ (the same region where the SEC energy condition is not satisfied) and positive for $y > R/\sqrt{2}$.

The above integral could be computed. We write down here only the final expression. 
  One finds that 
	 \begin{equation}
W_{K}(r) = -\frac{16R^{5}r^{3}}{m(2r^{2} + R^{2})^{3}}.
 \label{3.4}
 \end{equation}
The Komar energy is always negative, with	
 \begin{equation}
W_{K}(0) = 0,~~~~ W_{K}(R) = -\frac{16R^{2}}{27m},~~~~0\leq r \leq R.              
 \label{3.5}
 \end{equation}
Moreover, $W_{K}(r)$ has a minimum at $r = R/\sqrt{2}$, with $W_{K,min} = -R^{2}\sqrt{2}/m$. This is exactly the value where the SEC starts to be obeyed.
The negative value of $W_{K}(R)$ is probably a consequence of the negative pressure from (2.10). Inserting all fundamental constants in (3.5), we get for the Komar mass $W_{K}(R) = -16c^{4}R^{2}/27G^{2}m$. For a Solar mass star, one obtains $W_{K}(m_{\odot},R_{\odot}) \approx -10^{35}$ grams, with $|W_{K}(m_{\odot},R_{\odot})|$ close to the mass of the Sun, $10^{33}$ grams. If we consider, for example, a neutron star or a quark star, having the radius $R = 9m/4$ (a little bit more than its Schwarzschild radius), Eq. (3.5) gives us $W_{K}(m) = -3mc^{2}$. It is clear from here that $W_{K}$ is negative due to the interior negative pressure and negative radial acceleration from (2.11).

\section{Conclusions}
The Schwarzschild metric for a star interior is considered not to be realistic due to the constant energy density. Moreover, there is a singularity for some $r<(9/4)m$ and the pressure is divergent there.

 It is well known that the above Schwarzschild solution is the only static conformally flat solution with positive energy density and pressure. We argued in this paper that, using a negative pressure, a solution with constant energy density may be obtained for a conformally flat geometry. We further show that the pressure and the energy density are finite everywhere inside the object and the stress tensor of the inner fluid is investigated. The negative pressure leads to a repulsive gravitational field. The energy conditions for $T_{ab}$ are generally satisfied. The Komar mass resulting from the conformally flat metric is computed and discussed.

\end{document}